\documentclass[structabstract]{aa}  
\usepackage{natbib}
\bibpunct{(}{)}{;}{a}{}{,} % to follow the A&A style
\usepackage{graphicx}
\usepackage[varg]{txfonts}
\begin{document}
   \title{Testing the theory of colliding winds: The periastron passage\\
of 9 Sagittarii}
   \subtitle{II. Radio monitoring\thanks{Based on observations with the
              Karl G. Jansky Very Large Array (VLA), which is operated by
              the National Radio Astronomy Observatory.
              The National Radio Astronomy Observatory is a facility of
              the National Science Foundation operated under cooperative
              agreement by Associated Universities, Inc.}}

   \author{R. Blomme\inst{1}
           \and G. Rauw\inst{2}
           \and D. Volpi\inst{3}
           \and Y. Naz\'e\inst{2,}\thanks{F.R.S.-FNRS Senior Research Associate}
           \and R. Prinja\inst{4}
          }

   \offprints{R. Blomme}

   \institute{Royal Observatory of Belgium,
              Ringlaan 3, B-1180 Brussel, Belgium\\
              \email{Ronny.Blomme@oma.be}
             \and
              Groupe d’Astrophysique des Hautes Energies, STAR,
              Universit\'e de Li\`ege, 
              Quartier Agora (B5c, Institut d’Astrophysique et de
              G\'eophysique),
              All\'ee du 6 Ao\^ut, 19c,  B-4000 Sart Tilman, Li\`ege, Belgium
             \and
             \'{E}cole polytechnique de Bruxelles, Universit\'e Libre de Bruxelles, Avenue Franklin Roosevelt 50,
             B-1050 Brussels, Belgium
             \and
              Department of Physics \& Astronomy, University College London,
              Gower Street, London WC1E 6BT, UK
            }

   \date{Received date; accepted date}

% \abstract{}{}{}{}{} 
% 5 {} token are mandatory
 
  \abstract
  % context heading (optional)
  % {} leave it empty if necessary  
   {Colliding winds in massive-star binaries generate non-thermal radio emission and an increased X-ray emission.}
  % aims heading (mandatory)
   {We study the radio emission of the non-thermal emitter 9~Sgr, which is a long-period and highly eccentric binary.}
  % methods heading (mandatory)
   {We conducted a monitoring campaign of 9~Sgr around its 2013--2014 periastron passage, obtaining radio, X-ray, and optical
   spectroscopy data. In this paper, we analyse the radio data.}
  % results heading (mandatory)
   {The radio data at 3.6, 6, and 20 cm show a maximum near periastron. Other non-thermal radio emitters
   	among O-type binaries have a minimum flux near periastron, because the synchrotron emission is absorbed by
   	the free-free absorption in the two stellar winds. This is not the case for 9~Sgr,
   	because even at periastron the orbital separation is sufficient to maintain the wind-wind collision and its associated synchrotron emission outside the effective free-free radius of both stars. Our standard modelling with a local magnetic field that decreases inversely proportional to the distance
   from the stars ($1/r$) fails to explain the observed fluxes. Instead, we need to adopt a local magnetic field
   that decreases faster than $1/r$.}
  % conclusions heading (optional), leave it empty if necessary 
   {While our model provides an acceptably good fit to the data, there is clearly still room for improvement. More
   	advanced modelling should
   	include magnetohydrodynamics and more detailed physics of the
   	injection, acceleration, and cooling of the relativistic electrons. Radio observations at orbital phases away from periastron
   	can put further constraints on the models.}


   \maketitle
\nolinenumbers
%
%________________________________________________________________

\section{Introduction}

\citet{Bieging+89} made the first extensive radio survey of hot stars. Many of the stars detected show thermal emission
due to the free-free process in their ionized stellar winds. For these stars, the measured fluxes
can be used to derive the mass-loss rate \citep{Wright+Barlow75,Panagia+Felli75,Puls+08,RubioDiez+22}.

A number of stars, however, showed non-thermal radio emission. These were recognized by variability, or by a flux that is too
high to be explained by the free-free mechanism in the stellar wind. Over the years,
many more of these non-thermal radio emitters were found, as can be seen from the catalogue by 
\citet{DeBecker+Raucq13}\footnote{Updated version: \url{https://www.astro.uliege.be/~debecker/pacwb/}}.

The non-thermal radio emission among these hot stars turns out to be due to colliding-wind 
binaries. This was found for Wolf-Rayet stars \citep{Dougherty+Williams00},
as well as for O-type binaries \citep[e.g.][]{Blomme-HD167971+07,VanLoo08OB2no9,Blomme-Cyg8A+10,Kennedy-20No5,Benaglia15-93129}.
When the two winds in a massive-star binary collide, a shock is formed on either side of the
contact discontinuity.
These shocks accelerate a fraction of the electrons up to relativistic speeds, due to the first-order Fermi mechanism
\citep[also called diffusive shock acceleration,][]{Bell78,Reitberger+14,Pittard+21}.
 As these electrons spiral in the magnetic field, they emit synchrotron radiation that is seen as
non-thermal radio emission \citep{Eichler+Usov93}.

Synchrotron emission due to electrons accelerated in shocks occurs in other astrophysical contexts as well, 
such as supernova remnants or interplanetary shocks. Colliding-wind
binaries have a predictable variation in shock parameters, thereby providing a controlled experiment to test our knowledge of the acceleration process. They furthermore
provide the opportunity 
to study this in an environment where parameters such as the density, shock speed, and magnetic field
are quite different than those of supernova remnants or interplanetary shocks.
The colliding-wind binaries may also provide a handle on investigating the clumping or porosity
in stellar winds \citep{Pittard07}, and are thus relevant for determining the mass-loss rates \citep{Puls+08}.

In this paper, we study the radio fluxes of \object{9~Sgr} (HD~164794).
These were obtained during a monitoring campaign of its 2013--2014 periastron passage.
In parallel, X-ray data and optical spectra were also acquired. Their analysis has been presented
in \citet[][hereafter Paper I]{Rauw+16}.

After some initial indications of binarity \citep{Rauw+02,Rauw+05}, a spectroscopic orbit was finally
determined by \citet{Rauw+12}. In Paper I, we refined the orbital parameters, and determined the period to be $9.1 \pm 0.012$ year
and the eccentricity to be $e = 0.71 \pm 0.007$.
9~Sgr was also observed astrometrically \citep{Sana+14,LeBouquin+17,Fabry+21}. By
combining the spectroscopic and astrometric information, \citet{Fabry+21} made a full determination
of the binary orbit and the stellar parameters of the two components. Their value of the period is
$P = 8.9 \pm 0.2$~year and the eccentricity is $e = 0.649 \pm 0.009$.
The spectral types are O3V((f$^{+}$)) for the primary and O5V((f)) for the secondary.

\citet{Abbott+84} found 9~Sgr to be a non-thermal radio emitter based on its variable flux. It also showed a negative spectral
index\footnote{The spectral index, $\alpha$, is derived from the behaviour of the flux with frequency: $F_\nu \propto \nu^\alpha$.},
while a value of about $+0.6$ would be expected from the free-free mechanism.
The non-thermal nature was confirmed by further data collected by \citet{Bieging+89}.
The later observations by \citet{Rauw+02} also showed the negative spectral index.
\citet{Blomme+Volpi14} collected all VLA archive radio data and determined the radio light curve at 2, 3.6, 6, and 20 cm.
Most of the fluxes were much higher than expected from free-free emission of the stellar winds, and therefore
confirmed the non-thermal nature of the emission. Only the 2-cm data were of sufficient quality to show 
clear phase-locked variability with the orbit.

In Sect.~\ref{section observations} we present the 9~Sgr radio observations collected during the 2013--2014 monitoring campaign
and their data reduction. Sect.~\ref{section radio light curve} discusses the radio light curve.
In Sect.~\ref{section modelling} we present our theoretical model, and we apply it in Sect.~\ref{section results}.
In Sect.~\ref{section conclusions} we present our conclusions.

\section{Observations}
\label{section observations}

\subsection{VLA data}

We monitored the periastron passage of 9~Sgr with the Karl G. Jansky Very Large Array (VLA) for more than a year, 
covering the time from February 5 2013 to May 19 2014 (programmes 13A-160 and 14A-137).
We had 17 observing runs, spaced approximately uniformly during that
period (see Table~\ref{table radio data phase calibrator} 
for the observing log). During the monitoring campaign, the VLA configuration changed
from D (lowest angular resolution) to A (highest angular resolution).

A single observing run consists of
observations in the X-band (8.0--10.0 GHz, 3.6 cm), C-band (4.5--6.5 GHz, 6 cm), and L-band
(1.0--2.0 GHz, 20 cm).
For the X and C-band, 16 spectral windows of 128 MHz were used (64 MHz for the L band), 
where each spectral window consists of 64 channels of 2 MHz (1 MHz for the L band).
For the first four observations, no L-band data were collected, because
the high background due to the nearby Hourglass Nebula 
(at a distance of 2.9\arcmin\ from 9~Sgr) makes the detection of 9~Sgr impossible
when the VLA is in its lowest angular resolution (see also Sect.~\ref{section data reduction}).
Due to a technical problem, the C-band data of the first observation were not usable.

For each band we followed the observing sequence that starts with the phase calibrator then the target and then the
phase calibrator again, first in the C band then in the X band and finally in the L band.
The phase calibrator used is \object{J1820-2528}, at an angular distance of $\sim$~4\degr\ from 9~Sgr.
The measured phase calibrator fluxes are listed in Table~\ref{table radio data phase calibrator}.
At the end of each run, we observed in
all three bands the flux calibrator \object{3C286} = J1331+3030, which also serves as 
the bandpass calibrator.
Typical times on target are $\sim$~2 min in the X band and C band ($\sim$~3 min for the first four observations)
and $\sim$~2 min in the L band ($\sim$~3 min for the last six observations).

\subsection{Data reduction}
\label{section data reduction}

For the data reduction, we used 
CASA\footnote{\url{http://casa.nrao.edu}}
\citep[Common Astronomy Software Application,][]{CASATeam+22},
version 6.5.
While reading in the data, flags were automatically applied for 
focus problems, an incorrect subreflector position, an
off-source antenna position, and missing antennas. Further flagging was
done by visual inspection and is mainly concerned with removing the effect
of Radio Frequency Interference (RFI).

We started the calibration sequence by assigning the correct flux to the
flux calibrator 3C286. A model for this calibrator was used, as it is 
spatially resolved by the VLA. After delay and bandpass calibration,
the phase and amplitude were calibrated and the flux scale was transferred
to the phase calibrator J1820-2528. The calibrated data for the flux and phase calibrators were then inspected
visually for discrepant data. If these were found, they were flagged and the
calibration sequence was redone. The resulting fluxes of the phase calibrator
are listed in Table~\ref{table radio data phase calibrator}; they
show the slow flux variations that are typical of phase calibrators.

The target data were then calibrated
and an image was made. We ensured that the images cover the primary beam
and the pixel size was chosen so that it oversamples the synthesized beam
by a factor of at least 4 in both directions. 

Images obtained when the VLA was in a lower-angular-resolution configuration
(at the beginning of the monitoring campaign)
are dominated by the Galactic background and the \object{Hourglass Nebula}. 
The Hourglass Nebula is an H II region
that is being ionized by \object{Herschel 36}, a Trapezium-like multiple system \citep{Campillay+19}.
Sidelobes of this strong radio source interfere with the measurement
of the much weaker 9~Sgr.

To remove the effect of this background, we systematically dropped the
visibilities on the shortest baselines, thereby reducing the effect of the extended emission.
We used the CASA \texttt{tclean} procedure to clean the image around the Hourglass Nebula.
We used the multiscale deconvolver with \texttt{briggs} weighting (robustness parameter 0.5).
We applied multi-frequency synthesis, thereby combining all the frequency channels within a band into a single
continuum imaging.
The visibilities we obtained in this way were subtracted from the original visibility data, thereby
further reducing the effect of the extended emission. The resulting image was then cleaned further,
and we stopped the cleaning when the residual image no longer showed significant flux at the position of 9~Sgr.

For all our data, we also tried to apply a single iteration of self-calibration (phase only), but this did not always
improve the result. The data where self-calibration was successfully applied are indicated in
Table~\ref{table radio data 9 Sgr}.

We then applied the CASA task \texttt{imfit} to measure the fluxes of 9~Sgr. This fits a two-dimensional elliptical Gaussian
to 9~Sgr, from which we find the integrated flux, and its corresponding error bar. That error bar includes only the
effect of the noise on the flux. At the spatial resolution of our observations, 9~Sgr is a point source, so we verified that the
two-dimensional Gaussian did not find an extended source. 

To obtain a better handle on any systematic effects, we applied a jack-knife approach.
We systematically dropped one antenna, redid the imaging, again subtracted the Hourglass, and remeasured the fluxes.
We also systematically varied the depth of the cleaning, the range of visibilities we dropped, and the \texttt{briggs} robust
parameter, we applied a \texttt{uvtaper} to the data, and we processed stokes='RR' and 'LL' separately. The range from the smallest and largest
flux values from all these variant reductions then defines the final result. In Table~\ref{table radio data 9 Sgr}
we list the flux value (which is the middle of the range) and the error bar (which is half of the range).
The error bars also include 
the 5\% absolute calibration 
uncertainty\footnote{\url{https://science.nrao.edu/facilities/vla/docs/manuals/oss/performance/fdscale}}
(which was added in quadrature).

The earliest observations were taken when the VLA was in the D configuration (Table~\ref{table radio data phase calibrator}),
which has the lowest angular resolution. As a consequence, we were not able to detect 9~Sgr against the strong background
due to the Hourglass Nebula. In those cases, we measured the root-mean-squared flux around the position of 9~Sgr,
and assigned three times that value as the upper limit for that observation (Table~\ref{table radio data 9 Sgr}).

From the fluxes, we derived the spectral
index, $\alpha$, for the various combinations of bands. The error bar on $\alpha$ is derived from
standard error propagation, using the error bars on each of the
fluxes. When one of the fluxes has an upper limit, only a lower
limit on $\alpha$ can be determined. The spectral indices are listed in Table~\ref{table radio data 9 Sgr}.

\begin{figure}[ht]
	\centering
	\resizebox{\hsize}{!}{\includegraphics[bb=28 0 509 624]{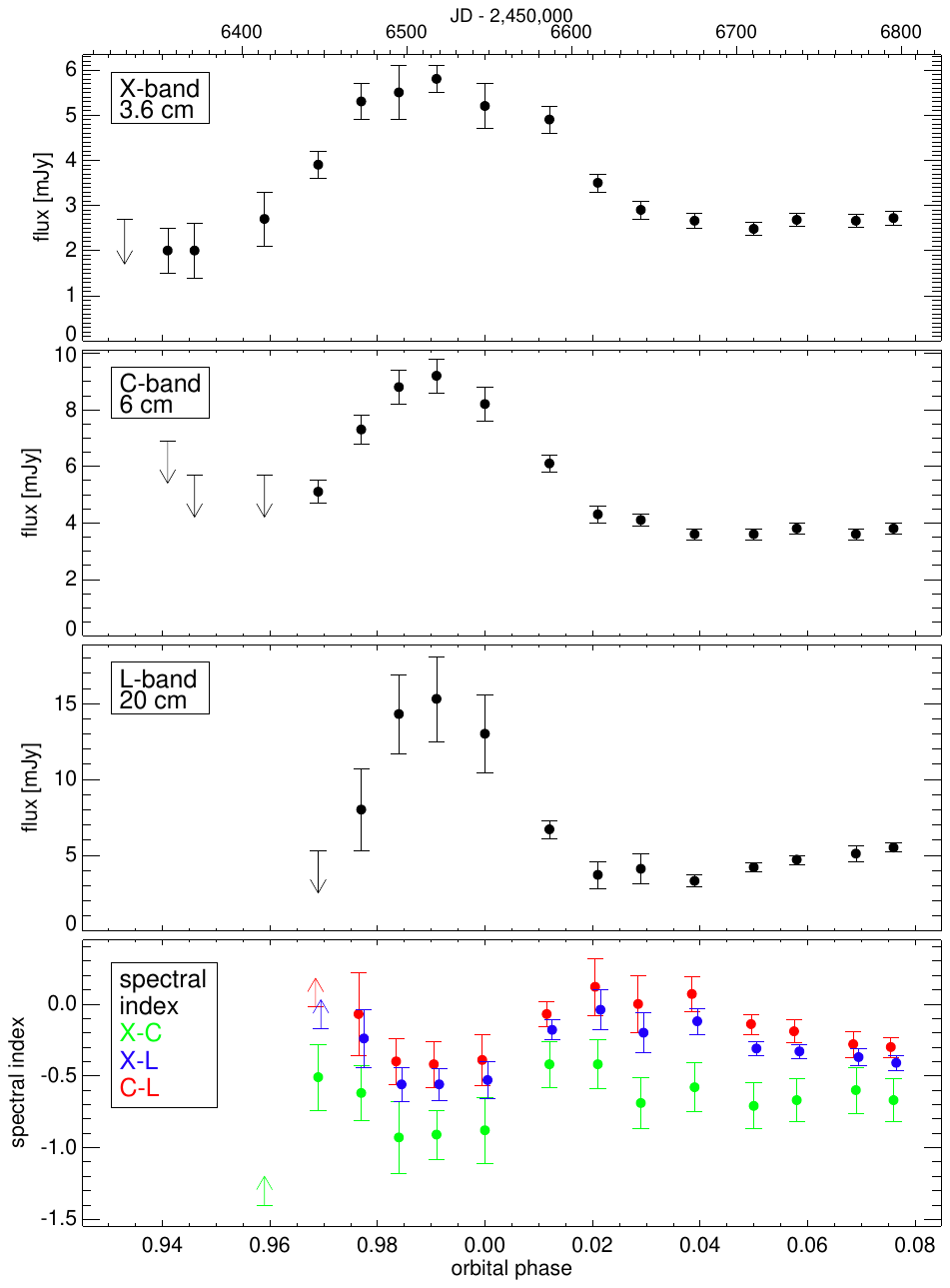}}
	\caption{Fluxes and spectral indices as a function of time. The top three panels show the radio fluxes and their error bars in the 
		X, C, and L bands, respectively. The bottom panel shows the spectral indices, colour-coded for the bands on which they are based.
		The symbols in the bottom panel are slightly offset in orbital phase to avoid them overlapping. The axis at the top shows the 
		heliocentric Julian date.}
	\label{fig fluxes}
\end{figure}

\section{Radio light curve}
\label{section radio light curve}

Figure~\ref{fig fluxes} shows the fluxes and spectral indices as a function of orbital phase. To determine
the orbital phase, we used the \citet{Fabry+21} ephemeris, as it is based on a combination of astrometric and
spectroscopic observation. Using the ephemeris from Paper I (which is based on spectroscopy only) would
shift all data in the plot by $0.002$ to the right.

The present data show a much cleaner radio light curve than the one shown in Fig.~1 of \citet{Blomme+Volpi14}. They
also cover the periastron passage much better.

The radio fluxes in all bands show clear variability. They reach their maximum just before periastron passage
(phase~$\approx$~0.99). There is no noticeable shift between the different bands in their time of maximum.
Shortly after periastron (phase~$\gtrsim$~0.03) the fluxes have decreased by a factor of 2--3 and level off,
except with a small turn-up at 20~cm.
The \citet{Blomme+Volpi14} Fig.~1 shows a further decrease at later orbital phases, but the fluxes do not seem
to fall below 1--2 mJy in all bands.

The bottom plot of Fig.~\ref{fig fluxes} shows the three possible spectral indices that can be determined.
All clearly show the presence of non-thermal radiation ($\alpha \lesssim 0.0$). The spectral indices become 
most negative just before periastron passage (phase 0.98--1.00). All this indicates that we are clearly
seeing the effect of the colliding-wind region.
The C-L indices are systematically smaller (in absolute value) than the X-C ones. We
discuss this further in Sect.~\ref{section results}.

The synchrotron emission is highest close to periastron. At first sight, this is consistent with
what could be expected. Near periastron in this highly eccentric binary, the stars will be closer
to one another, leading to a stronger collision (higher ram pressure) between their winds. In turn, this should lead to
a stronger synchrotron emission.

Nevertheless, observations of other colliding-wind binaries can show the opposite effect.
A minimum flux near periastron 
is seen in \object{HD~168112}, a colliding-wind binary with a 512-day period and an eccentricity of 0.75 \citep{Putkuri+23,Blomme_HD168112+24}.
The radio data in \citet{Blomme_HD168112+24} show that its fluxes are lowest near periastron, and are thermal.
\object{Cyg OB2\#9}, an 858-day period binary with an eccentricity of 0.71 
\citep{Naze+10,Naze+12,Blomme-Cyg9+13}, also has minimum flux near periastron
\citep{VanLoo+08,Blomme-Cyg9+13}.
Also the quadruple system \object{Cyg OB2\#5} has a flux minimum when the outer third component, in its $\sim$~6.7 yr period,
approaches the inner binary near periastron. \citet{Dzib+13} present Very Long Baseline Array observations of this system.
Away from periastron, the non-thermal emission of the colliding-wind region can be clearly resolved. As the system
approaches periastron, the emission disappears as it is absorbed in the combined stellar winds of the inner binary.
Among the Wolf-Rayet stars, \object{WR 140} also shows minimum radio flux near periastron
\citep[e.g.][]{Williams_WR140+90,White+Becker-WR140-95,Dougherty_WR140+05}. It
has a period of 7.94 yr and an eccentricity of 0.88 \citep{Marchenko+03}.

On the other hand, binaries that show behaviour similar to 9~Sgr, in the sense that they have their
maximum radio flux near periastron, are \object{Cyg OB2\#8A} and \object{HD~167971}.
Cyg OB2\#8A is a binary with a period of 21.9 days \citep{DeBecker+04,DeBecker+06} and
 an eccentricity of $e = 0.18 \pm 0.03$ \citep{Mossoux+20}.
It shows maximum radio emission near periastron \citep{Blomme-Cyg8A+10}.
HD 167971 is a triple system \citep{Leitherer+87}, where the radio fluxes are due to the interaction between the 
short-period inner binary and the third component. The radio emission is at maximum near the periastron passage
of the third component \citep{Blomme-HD167971+07,Ibanoglu+13,LeBouquin+17}. 
Among the Wolf-Rayet stars, \object{Apep} provides an example.  \citet{Marcote+21} present
	a Very Long Baseline Interferometer image where the synchrotron emission from the colliding-wind
	region between the two Wolf-Rayet stars is clearly resolved.
As Apep is the brightest and most
luminous non-thermal radio emitter among the colliding-wind binaries, this is a clear case in which most or all of the emitted synchrotron radiation reaches the observer.

The explanation for this different behaviour is the free-free absorption of the synchrotron photons by the
material in the stellar winds. When the two stars are closest together near periastron in an eccentric
binary, the two stellar winds can shield the emission from the colliding-wind region. Whether this happens
or not will depend on how close the stars are as seen projected on the sky, where the inclination of the
orbital plane plays an important role. Also the extent and density of the two stellar winds are relevant.
It is therefore not unexpected that some colliding-wind binaries have the minimum radio flux near periastron,
while others have their maximum flux. Specifically for 9~Sgr, \citet{Blomme+Volpi14} showed from a simple estimate
of the effective radius for radio emission that no significant stellar-wind absorption of the synchrotron emission
would occur. This does assume that the synchrotron emission arises from a limited region around the apex 
of the colliding-wind region. In reality, the synchrotron emission is generated in a much more extended region.

\begin{table}
	\caption{9~Sgr parameters used in modelling.}
	\label{table modelling parameters}
	\begin{center}
		\begin{tabular}{lcc}
			\hline\hline
			\noalign{\vskip 1mm}
			Parameter & \multicolumn{2}{c}{Value} \\
			& Primary & Secondary \\
			\hline
			\noalign{\vskip 1mm}
			$T$ (MJD)                  & \multicolumn{2}{c}{$ 56547       \pm   12$} \\
			$P$ (d)                  & \multicolumn{2}{c}{$3261        \pm   69$} \\
			$e$                       & \multicolumn{2}{c}{$0.648        \pm   0.009$} \\
			$\Omega$ ($\degr$)         & \multicolumn{2}{c}{$67.3         \pm   0.4$} \\
			$\omega$ ($\degr$)         & \multicolumn{2}{c}{$210.7         \pm   2.3$} \\
			$i$ ($\degr$)              & \multicolumn{2}{c}{$86.5         \pm   0.5$} \\
			distance (kpc)            & \multicolumn{2}{c}{$1.31          \pm   0.06$} \\
			$K$ (km\,s$^{-1}$) & $36^{+4}_{-1}$ & $49 \pm 3$\\
			\noalign{\vskip 1mm}
			$M$ (M$_{\sun}$) & $53^{+7}_{-6}$ & $39^{+6}_{-3}$ \\
			\noalign{\vskip 1mm}
			$a$ (R$_{\sun}$)      & $1770^{+200}_{-65}$ & $2410 \pm 160$ \\
			\noalign{\vskip 1mm}
			Spectral type            & 	O3V((f$^{+}$)) & O5V((f)) \\
			\noalign{\vskip 1mm}
			$T_{\rm eff}$ (K)              & $46000 \pm 1000$ & $42000 \pm 1000$ \\
			$\log$ $g$ (cgs)           & $3.87 \pm 0.20$  & $3.87 \pm 0.20$ \\
			$R$ (R$_\sun$)              & $10.8 \pm 1.0$ & $8.9 \pm 1.2$ \\
			$\log L/L_\sun$            & $5.68 \pm 0.08$ & $5.35 \pm 0.08$ \\
			$\log \dot{M}$ (M$_\sun$ yr$^{-1}$)     & $-6.6 \pm 0.2$ & $-6.6 \pm 0.2$ \\
			scaled $v_\infty$ (km\,s$^{-1}$)       & $2850$ & $2100$ \\
			\hline\\
		\end{tabular}
	\end{center}
	\tablefoot{Parameters derived by \citet{Fabry+21}, except for the scaled $v_\infty$ (see Sect.~\ref{section results}).
		The periastron argument, $\omega$, is from the astrometry. Its value is shifted by 180\degr\ compared to that 
		traditionally used in fitting the radial velocity curve.
	}
\end{table}

\section{Modelling}
\label{section modelling}

The model we use combines the synchrotron emission code of \citet{Blomme-Cyg8A+10} and the radiative transfer model
with the adaptive grid scheme of \citet{Blomme+Volpi14} and \citet{Blomme+17}.
The current model is an improvement over the one used by \citet{Blomme+Volpi14} for 9~Sgr, because it includes
much more physics of the synchrotron emission.
For the full details, we refer to 
\citet[][their Appendix A2--A6]{Blomme-Cyg8A+10}, which relies
heavily on the \citet{VanLoo} PhD thesis, where the citations to the original papers can be found.

Specifically for 9~Sgr, we have sufficient spectroscopic and astrometric information to allow us, for any given orbital phase, to put 
the two components in our three-dimensional simulation box. This box has its xy plane corresponding to the projection on the sky 
and the positive z axis pointing towards the observer. 
This configuration is different from that used in 
\citet[][their Appendix A.1, where the xy-plane contained the true orbit]{Blomme-Cyg8A+10}. The size of the box is 20\,000 $\mathrm{R}_\sun$. The origin is taken to be the centre of mass of the binary system.

No hydrodynamical model was used to determine the contact discontinuity and the shocks.
Instead we
used the \citet{Antokhin+04} equations to derive the position of the contact discontinuity in a two-dimensional plane containing the two stars, at each orbital phase.
We assumed symmetry along the line connecting the two stars, so that later on we simply rotated 
our results into three dimensions.
The \citeauthor{Antokhin+04} equations are appropriate for a colliding-wind region that is in the adiabatic regime
(as is the case for 9~Sgr -- see Sect.~\ref{section results}).
The radiative regime on the contrary would show a considerable amount of structure where the two winds collide, due to instabilities
\citep{Stevens+92}.

For the calculation of the synchrotron emission, we assumed that the two shocks
coincide with the contact discontinuity. We also assumed that the shocks are strong, i.e.
the compression ratio $\chi=4$. For a number of points on the two shocks, we generated a population of relativistic electrons.
The injection mechanism is not well known, so we cannot calculate ab initio how many relativistic electrons
are generated. Instead, we introduced the parameter, $\zeta$, which is the fraction of energy
that gets transferred from the shock to the relativistic electrons.
For $\zeta$ we took the value 0.05 \citep{Blandford+Eichler87,Eichler+Usov93}.
The momentum distribution of the electrons is a power law modified for the effect of inverse Compton cooling.
The exponent of the power law is set by the $\chi$ value that we have assumed.
Additionally, we needed to specify the lowest momentum that we considered, for which we took
$p_0 = 1.0$~MeV/$c$, with $c$ the speed of light.
We followed the change in the momentum distribution as the electrons moved away from the shock, and lost energy due
to advection and cooling. In this we adapted the time step as needed, decreasing it to reach the required precision,
or increasing it to reduce the computing time.

At each point on the tracks we followed in the two-dimensional plane, we could then determine the synchrotron emissivity. 
This is based on the
local momentum distribution of the relativistic electrons, and takes into account the Razin effect \citep{Ginzburg+Syrovatskii65}. 
This gives us an irregular, two-dimensional grid where the
synchrotron emissivity is specified.

In our standard model we take the \citet{Weber+Davis67} approach for the magnetic field. Further out in the stellar winds, 
the local magnetic field, $B(r)$, at distance $r$ from the centre of the star is given by
\begin{equation}
B(r) = B_* \frac{v_{\rm rot}}{v_\infty} \frac{R_*}{r},
\label{eq magnetic field}
\end{equation}
where $B_*$ is the surface magnetic field and $R_*$ is the radius of the star. 
For the ratio of the rotational velocity over the terminal wind velocity,
we assume the value $v_{\rm rot}/v_\infty = 0.1$.
No polar angle dependence is included in this. We stress, however, that inhomogeneities in the stellar
winds will lead to a local magnetic field that can be quite different from the approximation given here.
We neglect any amplification of the field \citep{Bell04} or any reduction by magnetic reconnection.
In Sect.~\ref{section results} we show that the observations indeed require a different radius dependence.

To determine the emergent intensity towards the observer, we now return to the three-dimensional grid. For a given (x,y) point,
we can solve the radiative transfer equation for a ray going from $z=-\infty$ to $z=+\infty$. The synchrotron emissivity
at any point in the three-dimensional grid 
can be found by rotating the three-dimensional point into the two-dimensional plane and interpolating. 
Free-free opacity and emissivity of the stellar-wind material are obtained by analytical
calculation. 
We assume that the material in the stellar winds is at half the effective temperature of the star.
No clumping or porosity is included, but we note that the \citet{Fabry+21} mass-loss rates are derived with clumping
included. We also neglect any free-free contribution from the colliding-wind region itself.

\begin{figure}[b]
	\centering
	\resizebox{\hsize}{!}{\includegraphics[bb=48  0 509 480]{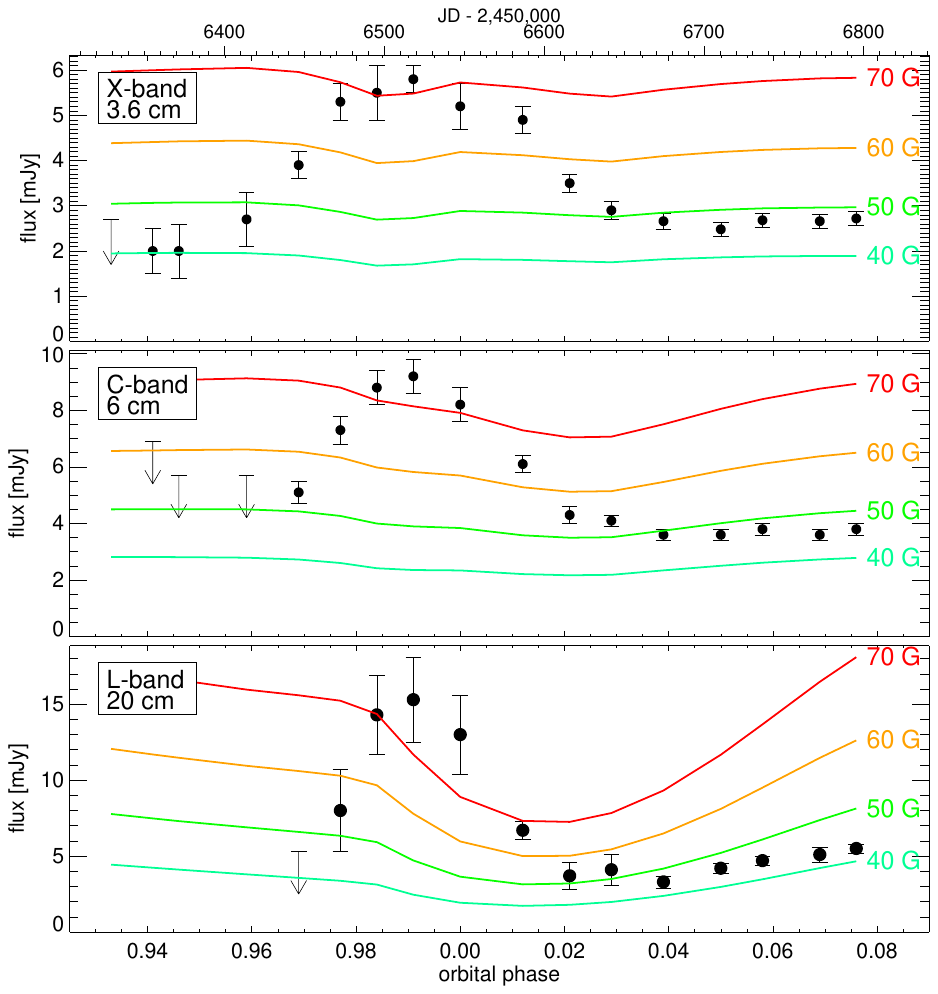}}
	\caption{Theoretical fluxes of our standard model for different values of the magnetic field ($B_*$), overplotted on the observed data.
		The symbols are as in Fig.~\ref{fig fluxes}.}
	\label{fig Bsurf}
\end{figure}

\begin{figure}[ht]
	\centering
	\resizebox{\hsize}{!}{\includegraphics[bb=30 30 310 190]{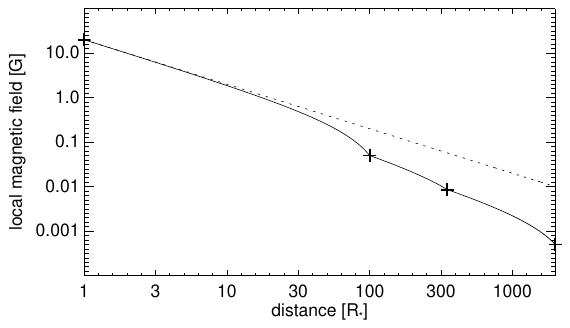}}
	\caption{Local magnetic field as a function of distance from the centre of the star. 
		The full line shows the step-wise $B_*$ approach we 
	introduced in Sect.~\ref{section results}
	to describe the magnetic field. It corresponds to the magnetic field law used to obtain the best
	fit to the observed fluxes (Fig.~\ref{fig result}). For comparison, the dotted line shows the magnetic field for 
a single fixed $B_*$, as in our standard model.}
	\label{fig magnetic}
\end{figure}

\begin{figure}[ht]
	\centering
	\resizebox{\hsize}{!}{\includegraphics[bb=28 0 509 480]{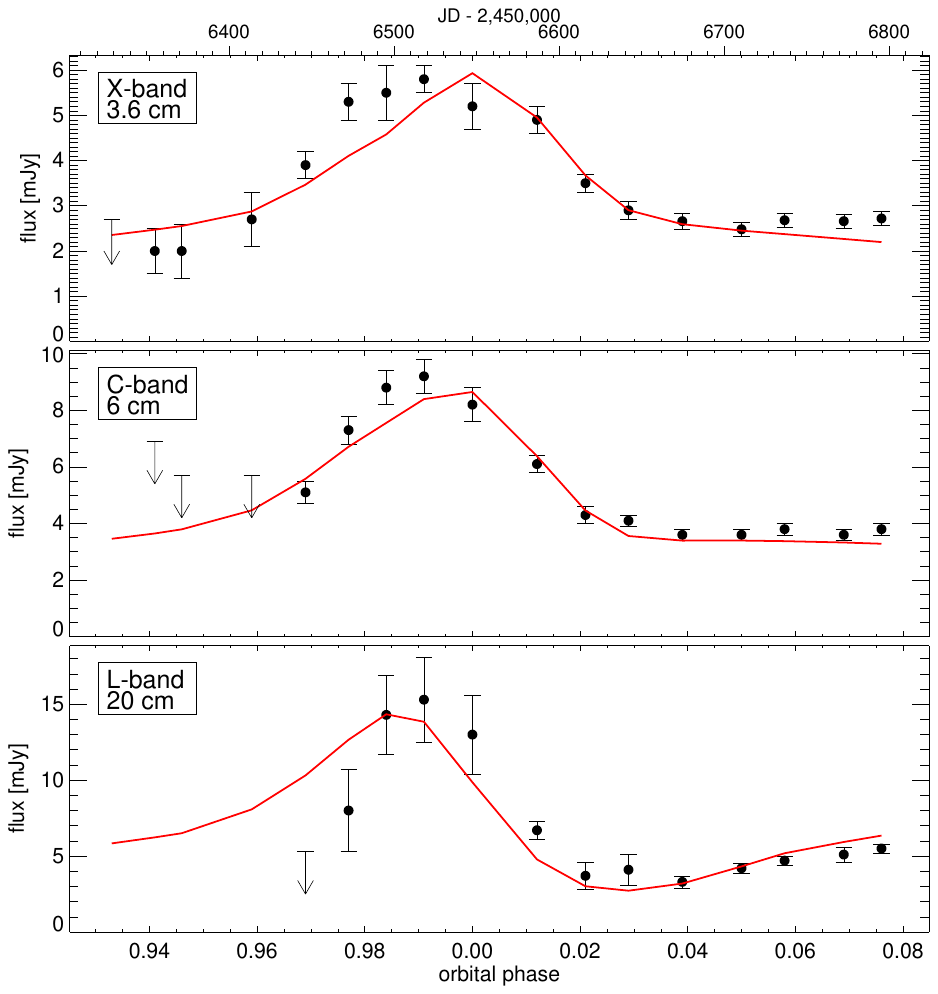}}
	\caption{Best fit of our model to the observed fluxes. The model is based on the local magnetic field law
	as shown in Fig.~\ref{fig magnetic}.}
	\label{fig result}
\end{figure}

In the radiative transfer procedure, we used an adaptive grid scheme. We started by solving the emergent intensity required
in the flux calculation in $256$
gridpoints in the z direction, and refined the grid where needed to get the required precision.
The flux was obtained by integrating over the emergent intensity. This integration started with a $256\times 256$ grid,
and each cell was refined to obtain the required precision. The resulting value was scaled for the distance to obtain the 
flux received at Earth.

\section{Results and discussion}
\label{section results}

The parameters we used in our modelling are listed in Table~\ref{table modelling parameters}.
For the orbital parameters, \citet{LeBouquin+17} find a tension 
between their astrometric data 
and the spectroscopic orbit determined in Paper I.
This is resolved by the analysis of the combined astrometric and spectroscopic data
by \citet{Fabry+21}, so we use their values here.

We also used their spectroscopic stellar atmosphere parameters and mass-loss rates based on their {\sc Fastwind} modelling.
They determine the clumping factor, but for simplicity, our model assumes smooth winds.
The terminal velocity was derived from UV P Cygni profiles.
\citet{Prinja+90} measured a value of  
$v_\infty = 2750$~km\,s$^{-1}$ and \citet{Groenewegen+89} had $v_\infty = 2950 \pm 150$~km\,s$^{-1}$, but these values
do not distinguish between the two components. The theoretical predictions of \citet{Muijres+12} give
3800 and 2800 km\,s$^{-1}$ for these spectral types. Their Fig.~9 suggests that a simple scaling can
be applied to make them agree with the observed $v_\infty$. We assumed that the 
primary component dominates the UV P Cygni profiles and we therefore assigned the average value of the \citeauthor{Prinja+90} and \citeauthor{Groenewegen+89} determinations to the primary.

An important assumption of our model is that the colliding-wind region is in the adiabatic regime
(Sect.~\ref{section modelling}).
With the parameters from Table~\ref{table modelling parameters}, we can now verify that this is the case.
We compared the cooling time ($t_{\rm cool}$) to the escape time ($t_{\rm esc}$), using the 
equation from \citet{Stevens+92}:
\begin{equation}
\psi = \frac{t_{\rm cool}}{t_{\rm esc}} \approx \frac{v_8^4 d_{12}}{\dot{M}_{-7}},
\end{equation}
where $v_8$ is the wind velocity in units of 1000~km\,s$^{-1}$, $d_{12}$ the distance to the contact discontinuity
in units of $10^7$ km, and $\dot{M}_{-7}$ the mass-loss rate in units of $10^{-7} {\rm M}_\sun {\rm yr}^{-1}$.
At periastron, we find $\psi \approx 1500$ for primary and $\psi \approx 370$ to the secondary. These high $\psi$ values show that 
the collision region is indeed adiabatic, even when the two stars are closest to one another.

In trying to fit the observations with our model, we started by exploring the effect of the magnetic field.
Figure~\ref{fig Bsurf} compares the theoretical fluxes of our standard model to the observed data. The range
of surface magnetic field we chose\footnote{Our assumed values of $B_*$ fall well below the upper dipolar field strength
	limit of 605 G determined by \citet{Neiner+15}.}
($B_* = 40 - 70$~G) covers the range of observed fluxes in all three bands,
but clearly fails to explain their dependence on orbital phase. At 6 and 20 cm, the theoretical fluxes show a
minimum just after periastron passage (phase $\approx 0.02$). As pointed out in  Sect.~\ref{section radio light curve},
this behaviour is seen in a number of colliding-wind binaries, but clearly does not apply to 9~Sgr.

We explored further the effect of the various parameters of the synchrotron emission mechanism
by varying the shock strength, $\chi$,
the fraction of shock energy that gets transferred to the relativistic electrons ($\zeta$), and the low momentum cutoff, $p_0$.
The effect of changing any of these values is roughly equivalent to a global scaling of the fluxes. The minimum just
after periastron passage remains and the theoretical model therefore does not explain the observations. Similarly,
changing the mass-loss rate and terminal velocity of the winds does not provide a good fit.

In exploring the magnetic field (Fig.~\ref{fig Bsurf}), we saw that no single value could explain the observations.
However, different values can explain the fluxes in limited ranges of orbital phase. Specifically, fluxes very close to
periastron require a higher magnetic field than fluxes further away in orbital phase. This leads to the idea
of using a local magnetic field law that decreases faster with radius than given in Eq.~(\ref{eq magnetic field}).
A similar approach was taken by \citet{Benaglia+25} to explain the radio data of HD 93129A.
From a more theoretical point of view, we note that at the apex the shocks will be perpendicular, which allows
the magnetic field to be amplified. Further out, the shocks are quasi-parallel, so almost no amplification
occurs \citep{Pittard+20}. This is consistent with the idea of a magnetic field decreasing faster than $1/r$.

We therefore introduced a step-wise $B_*$ as a function of radius as shown in Fig.~\ref{fig magnetic}. In this, $B_*$ then loses its meaning as the surface magnetic field, but still serves as the scaling factor in Eq.~(\ref{eq magnetic field}). 
We used four breakpoints in this magnetic field law, as indicated by the crosses in Fig.~\ref{fig magnetic}.
The figure also shows how the magnetic field declines faster than the $1/r$ dependence of the standard approach. 
We then explored a large range of values for the position and value for the break points. The best fit we obtained (as judged by eye)
is shown in Fig.~\ref{fig result}.

We now obtained a good fit that follows the flux changes with orbital phase very well.
The phase of maximum flux for each band is very close to the observed one. At later phases, the theoretical
fluxes continue to follow the observed ones. Even the small turn-up at later phases for 20 cm is present in the model.

Nevertheless, some discrepancies remain. Somewhat before periastron (phase $\approx 0.98$), our model underestimates
the 3.6 and 6 cm fluxes, but overestimates the 20 cm ones. We tried to remedy this situation by applying changes to the
mass-loss rates or terminal velocities of both stars, but found no significant improvement. We also experimented with the
various parameters of the synchrotron emission ($\chi$, $\zeta$, and $p_0$), but these did not improve the result either.
We note that the largest discrepancy at 20~cm occurs for data where the background due to the
Hourglass Nebula is strongest (Sect.~\ref{section data reduction}). The upper limit and the uncertainty on the measured flux
may in those cases have been set too sharp.

In Sect.~\ref{section radio light curve} we noted that the C-L spectral indices are systematically smaller 
(in absolute value) than the X-C ones. This 
could be indicative of the start of a turn-over in the radio spectrum, which may suggest the Razin effect
or synchrotron self-absorption. However, further experimenting with our model shows a more likely reason. When switching 
off the absorption in the stellar winds, we find that both spectral indices are nearly the same (with a value 
of $\alpha \approx -1.0$). Although most of the synchrotron emission is getting through, there is some free-free
absorption at 20~cm (where the effective free-free radius is largest), leading to a lower C-L spectral index.

We also checked if our best-fit model is consistent with the data presented in \citet{Blomme+Volpi14}. Beyond the 
orbital phases shown in Fig.~\ref{fig result}, 
the theoretical model flattens out. It reaches 1.0~mJy at 3.6~cm, 1.7~mJy at 6~cm, and 4.5~mJy at 20~cm at apastron.
 The 20~cm observation near phase $\sim$~0.6
in Fig.~1 of \citet{Blomme+Volpi14} is considerably higher, but we
note that \citet{Rauw+02} listed a much lower value for this observation. The difference is attributed to the difficulty of correcting
for the background due to the Hourglass Nebula.
The better-quality data of \citeauthor{Blomme+Volpi14} at 3.6 and 6 cm tend to be somewhat higher than the
theoretical values, by a factor of $\sim$~2. This indicates that further adjustments to the magnetic field law would be needed
if we wanted to explain any future high-quality radio data outside periastron.

In Paper I, we studied the X-ray emission during the periastron passage. The soft part of the X-rays could be shown to
be due to intrinsic shocks in the stellar winds, and is therefore not related to the colliding-wind region. The 
hard X-ray emission, on the other hand, varies with orbital phase, showing that it is formed in
the colliding-wind region. The colliding-wind region is clearly adiabatic and in that case a $1/D$ dependence is expected for the
X-ray emission, where $D$ is the separation between the two stars \citep{Stevens+92}. 
However, the observations do not agree with that. In Paper I, we suggest that this may be due to
a phase-dependent efficiency of particle acceleration. In the present paper, we introduce a magnetic field law
that falls off faster than $1/r$ to explain the radio data.
While these are two different explanations, it may well be that there is
a single underlying explanation for the discrepancies in both the X-rays and the radio observations.
One such possibility that was mentioned in Paper~I is shock modification \citep{Pittard+Dougherty06}. 
The relativistic electrons 
can diffuse upstream and create a shock precursor. The stronger magnetic
field at the apex would create more of these electrons, and therefore a stronger precursor.
As the shock itself is less strong, this would also lead to a reduction in the hard X-ray emission. 
In any case, it makes clear that our current modelling of these colliding winds is still incomplete. 

Improvements in this situation can come from detailed magnetohydrodynamical modelling that includes clumping and the evolution
of the local magnetic field. This would also correctly position the shocks with respect to the contact discontinuity.
The shocks themselves will be influenced by the back-reaction of the relativistic pressure.
Further improvements would come from quantitative constraints on the injection mechanism of the relativistic electrons.
And other acceleration processes could be included, 
such as magnetic reconnection, as well as other cooling mechanisms (synchrotron, relativistic Brehmstrahlung).
Some of these effects have already been included in modelling of colliding-wind binaries \citep{Pittard+21}.

\section{Conclusions}
\label{section conclusions}

We analysed the 9~Sgr radio data obtained with the VLA during the monitoring campaign of its 2013--2014 periastron passage.
Contrary to some other colliding-wind binaries with non-thermal radio emission, the fluxes of
9~Sgr show a maximum near periastron. With a period of 8.9 year, this system is therefore wide enough
that even at periastron only a limited fraction of the synchrotron radiation is absorbed by the free-free absorption
in the two stellar winds.

The reduction of the data at 20 cm obtained in the lower angular resolution VLA configurations is hampered by the large
background due to the Hourglass Nebula. But, overall, the flux values obtained have small error bars.
These data therefore present a substantial improvement over the 9~Sgr data that were previously
available \citep{Blomme+Volpi14}.

The high quality of the new data presents substantial challenges to our theoretical modelling.
Our standard approach where we assume the magnetic field law to decrease as $1/r$ from the star turns out to
be unable to fit the observations. Only by introducing a magnetic field law that falls off faster than $1/r$ can we
obtain an acceptable fit to the data. Even then, some discrepancies remain. While at least those at 20 cm
may be due to the data reduction problems, the other ones at 3.6 and 6 cm indicate a lack of sufficiently
detailed physics in our model.

The modelling can be improved in various ways, by including magnetohydrodynamics and more detailed physics of the
injection, acceleration, and cooling of the relativistic electrons. Observationally,
it would be interesting to obtain similar high-quality radio data at orbital phases away from periastron.
This would allow us see how well the model explains the behaviour of 9~Sgr radio emission over its complete orbit.

\begin{acknowledgements}
We thank
Joan Vandekerckhove for his help with the reduction of the
VLA data. 
This research has made use of the SIMBAD database, operated at CDS,
Strasbourg, France and NASA's Astrophysics Data System Abstract Service.
This work
has used the following software products:
CASA \citep{CASATeam+22}, and
IDL (\url{https://www.nv5geospatialsoftware.com/Products/IDL}).
We thank the National Radio Astronomy
Observatory (NRAO) for carrying out the Karl G. Jansky Very Large Array
(VLA) observations.
We thank the referee for comments that helped clarify the paper.
\end{acknowledgements}

\bibliographystyle{aa}
\bibliography{aa61771-26}

\begin{appendix}

\onecolumn
\section{Tables}

\begin{table*}[ht]
	\caption{Observing log of the VLA data and radio fluxes of the phase calibrator.}
	\label{table radio data phase calibrator}
	\centering
	\begin{tabular}{rrlccc}
		\hline\hline
		\noalign{\vskip 1mm}
		\multicolumn{1}{c}{date}  & \multicolumn{1}{c}{HJD} & \multicolumn{1}{c}{config} & \multicolumn{3}{c}{Fluxes phase calibrator} \\
		\cline{4-6}
		\noalign{\vskip 1mm}
		(yyyy-mm-dd) & -2\,450\,000 & & X-band (3.6 cm) & C-band (6 cm) & L-band (20 cm) \\
		&              & & (Jy) & (Jy) & (Jy) \\
		\hline
		\noalign{\vskip 1mm}
		2013-02-05 & 6329.05 &    D &  0.9701 $\pm$  0.0011 &                   --- &                   --- \\ 
		2013-03-04 & 6355.96 &    D &  0.8415 $\pm$  0.0021 &  0.9602 $\pm$  0.0031 &                   --- \\ 
		2013-03-21 & 6373.01 &    D &  0.9638 $\pm$  0.0005 &  1.0988 $\pm$  0.0011 &                   --- \\ 
		2013-05-01 & 6413.85 &    D &  0.9149 $\pm$  0.0005 &  0.9233 $\pm$  0.0010 &                   --- \\ 
		2013-06-02 & 6445.78 &   CD &  0.8556 $\pm$  0.0003 &  0.8853 $\pm$  0.0007 &  0.916 $\pm$  0.004 \\ 
		2013-06-28 & 6471.71 &    C &  0.9429 $\pm$  0.0006 &  0.9035 $\pm$  0.0003 &  0.909 $\pm$  0.004 \\ 
		2013-07-23 & 6496.70 &    C &  0.8320 $\pm$  0.0003 &  0.8005 $\pm$  0.0006 &  0.901 $\pm$  0.004 \\ 
		2013-08-15 & 6519.63 &    C &  0.8340 $\pm$  0.0002 &  0.8108 $\pm$  0.0010 &  0.897 $\pm$  0.004 \\ 
		2013-09-11 & 6546.57 &  CnB &  0.8301 $\pm$  0.0004 &  0.8042 $\pm$  0.0006 &  0.881 $\pm$  0.003 \\ 
		2013-10-19 & 6585.43 &    B &  0.8798 $\pm$  0.0004 &  0.8425 $\pm$  0.0006 &  0.864 $\pm$  0.004 \\ 
		2013-11-17 & 6614.36 &    B &  0.8138 $\pm$  0.0002 &  0.8298 $\pm$  0.0005 &  0.857 $\pm$  0.003 \\ 
		2013-12-15 & 6642.27 &    B &  0.8192 $\pm$  0.0006 &  0.7598 $\pm$  0.0007 &  0.799 $\pm$  0.005 \\ 
		2014-01-17 & 6675.19 &    B &  0.9092 $\pm$  0.0008 &  0.7955 $\pm$  0.0008 &  0.850 $\pm$  0.005 \\ 
		2014-02-20 & 6709.14 &    A &  0.8262 $\pm$  0.0007 &  0.7838 $\pm$  0.0003 &  0.816 $\pm$  0.003 \\ 
		2014-03-19 & 6736.03 &    A &  0.8306 $\pm$  0.0005 &  0.7553 $\pm$  0.0003 &  0.818 $\pm$  0.002 \\ 
		2014-04-23 & 6770.95 &    A &  0.8281 $\pm$  0.0007 &  0.8049 $\pm$  0.0006 &  0.822 $\pm$  0.001 \\ 
		2014-05-19 & 6796.90 &    A &  0.7997 $\pm$  0.0003 &  0.8867 $\pm$  0.0006 &  0.830 $\pm$  0.002 \\ 
		\hline
	\end{tabular}
	\tablefoot{We list observing date, heliocentric Julian date
		and configuration of the VLA. The fluxes of the phase calibrator
		J1820-2528 are listed next (at 3.6, 6, and 20~cm).
		All fluxes have been calibrated on the flux calibrator 3C286.
	}
\end{table*}

\begin{table*}
\caption{Radio fluxes and spectral indices
of the VLA data on 9~Sgr.}
\label{table radio data 9 Sgr}
\centering
\begin{tabular}{ccr@{ $\pm$ }lr@{ $\pm$ }lr@{ $\pm$ }lcr@{ $\pm$ }lr@{ $\pm$ }lr@{ $\pm$ }lcccccccccccccc}
\hline\hline
\noalign{\vskip 1mm}
date  & orbital & \multicolumn{6}{c}{9~Sgr flux} & & \multicolumn{6}{c}{9~Sgr spectral index} \\
\cline{3-8}
\cline{10-15}
\noalign{\vskip 1mm}
      (yyyy-mm-dd) & phase & \multicolumn{2}{c}{X-band}   & \multicolumn{2}{c}{C-band} & \multicolumn{2}{c}{L-band}  & & \multicolumn{2}{c}{sp index} & \multicolumn{2}{c}{sp index} & \multicolumn{2}{c}{sp index} \\
      &       & \multicolumn{2}{c}{(3.6 cm)} & \multicolumn{2}{c}{(6 cm)} & \multicolumn{2}{c}{(20 cm)} & & \multicolumn{2}{c}{X-C}      & \multicolumn{2}{c}{X-L}      & \multicolumn{2}{c}{C-L}    \\
      &       & \multicolumn{2}{c}{(mJy)}    & \multicolumn{2}{c}{(mJy)}  & \multicolumn{2}{c}{(mJy)}   & & \multicolumn{2}{c}{   }      & \multicolumn{2}{c}{   }      & \multicolumn{2}{c}{   }    \\
 \hline
 \noalign{\vskip 1mm}
 2013-02-05 &  0.933 & \multicolumn{2}{c}{$<$ 2.7} & \multicolumn{2}{c}{--} & \multicolumn{2}{c}{--} & & \multicolumn{2}{c}{--} & \multicolumn{2}{c}{--} & \multicolumn{2}{c}{--} \\
 2013-03-04 &  0.941 &  2.0  & 0.5    & \multicolumn{2}{c}{$<$ 6.9} & \multicolumn{2}{c}{--} & & \multicolumn{2}{c}{$>$ $-$2.4} & \multicolumn{2}{c}{--} & \multicolumn{2}{c}{--} \\
 2013-03-21 &  0.946 &  2.0  & 0.6    & \multicolumn{2}{c}{$<$ 5.7} & \multicolumn{2}{c}{--} & & \multicolumn{2}{c}{$>$ $-$2.0} & \multicolumn{2}{c}{--} & \multicolumn{2}{c}{--} \\
 2013-05-01 &  0.959 &  2.7  & 0.6    & \multicolumn{2}{c}{$<$ 5.7} & \multicolumn{2}{c}{--} & & \multicolumn{2}{c}{$>$ $-$1.4} & \multicolumn{2}{c}{--} & \multicolumn{2}{c}{--} \\
 2013-06-02 &  0.969 &  3.9  & 0.3    &  5.1  & 0.4    & \multicolumn{2}{c}{$<$ 5.3} & & $-$0.51 & 0.23  & \multicolumn{2}{c}{$>$ $-$0.17} & \multicolumn{2}{c}{$>$ $-$0.02}  \\
 2013-06-28 &  0.977 &  5.3  & 0.4  S &  7.3  & 0.5    &  8.0 & 2.7   & & $-$0.62 & 0.19  & $-$0.24 & 0.20  & $-$0.07 & 0.29  \\
 2013-07-23 &  0.984 &  5.5  & 0.6  S &  8.8  & 0.6  S & 14.3 & 2.6   & & $-$0.93 & 0.25  & $-$0.56 & 0.12  & $-$0.40 & 0.16  \\
 2013-08-15 &  0.991 &  5.8  & 0.3    &  9.2  & 0.6  S & 15.3 & 2.8 S & & $-$0.91 & 0.17  & $-$0.56 & 0.11  & $-$0.42 & 0.16  \\
 2013-09-11 &  0.000 &  5.2  & 0.5  S &  8.2  & 0.6  S & 13.0 & 2.6   & & $-$0.88 & 0.23  & $-$0.53 & 0.13  & $-$0.39 & 0.18  \\
 2013-10-19 &  0.012 &  4.9  & 0.3  S &  6.1  & 0.3    &  6.7 & 0.6   & & $-$0.42 & 0.16  & $-$0.18 & 0.07  & $-$0.07 & 0.09  \\
 2013-11-17 &  0.021 &  3.5  & 0.2    &  4.3  & 0.3    &  3.7 & 0.9   & & $-$0.42 & 0.17  & $-$0.04 & 0.14  & $+$0.12 & 0.20  \\
 2013-12-15 &  0.029 &  2.9  & 0.2    &  4.1  & 0.2    &  4.1 & 1.0   & & $-$0.69 & 0.18  & $-$0.20 & 0.14  & $+$0.00 & 0.20  \\
 2014-01-17 &  0.039 &  2.66 & 0.17   &  3.6  & 0.2    &  3.3 & 0.4   & & $-$0.58 & 0.17  & $-$0.12 & 0.09  & $+$0.07 & 0.12  \\
 2014-02-20 &  0.050 &  2.48 & 0.15   &  3.6  & 0.2    &  4.2 & 0.3   & & $-$0.71 & 0.16  & $-$0.31 & 0.05  & $-$0.14 & 0.07  \\
 2014-03-19 &  0.058 &  2.68 & 0.15   &  3.8  & 0.2    &  4.7 & 0.3   & & $-$0.67 & 0.15  & $-$0.33 & 0.05  & $-$0.19 & 0.08  \\
 2014-04-23 &  0.069 &  2.66 & 0.15 S &  3.6  & 0.2  S &  5.1 & 0.5   & & $-$0.60 & 0.16  & $-$0.37 & 0.06  & $-$0.28 & 0.09  \\
 2014-05-19 &  0.076 &  2.72 & 0.15 S &  3.8  & 0.2    &  5.5 & 0.3   & & $-$0.67 & 0.15  & $-$0.41 & 0.05  & $-$0.30 & 0.07  \\
\hline
\end{tabular}
\tablefoot{We list observing date and orbital phase according to the \citet{Fabry+21} ephemeris.
The fluxes of 9~Sgr
are listed next (at the X, C, and L band) and then the three spectral indices.
The 'S' notation next to a flux indicates that self-calibration was applied.
}
\end{table*}

\end{appendix}

\end{document}